\documentclass[preprint,12pt,fleqn]{elsarticle}
\usepackage{amssymb}
\usepackage{amsfonts}
\usepackage[english]{babel}
\usepackage{euscript}
\usepackage{bbm}
\usepackage{xfrac}

\journal{Journal of Geometry and Physics}

\newcommand{\fl}{\hspace*{-\mathindent}}
\newcommand{\textfrac}[2]{\textstyle{\frac{#1}{#2}}}

\newcounter{example_counter}

\begin{document}

\begin{frontmatter}

\title{
Deformations of infinite-dimensional Lie algebras,
\\
exotic cohomology, and integrable nonlinear
\\
partial differential equations
}

\author{Oleg I. Morozov}
\ead{morozov{\symbol{64}}agh.edu.pl}
\address{Faculty of Applied
  Mathematics, AGH University of Science and Technology,
  \\
  Al. Mickiewicza 30,
  Cracow 30-059, Poland}

\begin{abstract}
The important unsolved problem in theory of integrable systems is to find conditions guaranteeing existence
of a Lax representation for a given {\sc pde}. The exotic cohomology of the symmetry algebras  opens a way to
formulate such conditions in internal terms of the {\sc pde}s under the study. In this paper we consider certain
examples of infinite-dimensional Lie algebras with nontrivial second exotic cohomology groups and show that the
Maurer--Cartan forms of the associated extensions of these Lie algebras generate Lax representations for
integrable systems, both known and new ones.

\end{abstract}


\end{frontmatter}

% \linenumbers
%------------------------------------------------------------------

%\maketitle

%%%%%%%%%%%%%%%%%%%%%%%%%%%%%%%%%%%%%%%%%%%%%%%%%%%%%%%%%%%%%%%%%%%%%%%%%%%%%%%%%%%%%%%%%%%%%%%%%%%%%%

\section{Introduction}
The existence of a Lax representation is the key property of integrable equations,
\cite{Zakharov1991,LNM810}, and a starting setting for a number of techniques to study nonlinear partial
differential equations ({\sc pde}s) such as B\"acklund transformations, nonlocal symmetries and conservation
laws, recursion operators, Darboux transformations, etc. Although these structures are of great significance
in the theory of integrable {\sc pde}s, up to now the problem of finding conditions for a {\sc pde} to admit a
Lax representation is open. In \cite {Morozov2017} we propose an approach for solving this problem in
internal terms of the {\sc pde} under the study. We show there that for some {\sc pde}s their Lax
representations can be derived from the second
exotic\footnote{Unlike in \cite{Morozov2017}, in this paper we
follow \cite{Novikov2002} and use the term ``exotic cohomology'' instead of ``deformed cohomology'', since here we
discuss deformations of Lie algebras which are not related to ``deformed cohomology'' in the sense of
\cite{Morozov2017}.}
cohomology of the symmetry pseudogroups of the {\sc pde}s. The main advantage of this approach is that it allows
one to get rid of apriori assumptions about the defining equations of the Lax representation. In this paper we
generalize the constructions of \cite{Morozov2017}. We consider a deformation of the tensor product of the Lie
algebra of vector fields on a line and the algebra of truncated polynomials as well as certain extensions of
this deformation and show that at some values of the deformation parameter the Maurer--Cartan forms of the
obtained Lie algebras produce Lax representations for some known as well as some new integrable systems.

%%%%%%%%%%%%%%%%%%%%%%%%%%%%%%%%%%%%%%%%%%%%%%%%%%%%%%%%%%%%%%%%%%%%%%%%%%%%%%%%%%%%%%%%%%%%%%%%%%%%%%

\section{Preliminaries}\label{Preliminaries_section}

All considerations in this paper are local. All functions are assumed to be real-analytic.

\subsection{Coverings of PDEs}

The coherent geometric formulation of Lax representations, Wahlquist--Estabrook prolongation structures,
B\"acklund transformations, recursion operators, nonlocal symmetries, and nonlocal conservation laws is
based on the concept of differential covering of a {\sc pde}
\cite{KrasilshchikVinogradov1984,KrasilshchikVinogradov1989}.
In this subsection we closely follow
\cite{KrasilshchikVerbovetsky2011,KrasilshchikVerbovetskyVitolo2012} to present the basic notions of the theory
of differential coverings.

Let $\pi \colon \mathbb{R}^n \times \mathbb{R}^m \rightarrow \mathbb{R}^n$,
$\pi \colon (x^1, \dots, x^n, u^1, \dots, u^m)$    $\mapsto (x^1, \dots, x^n)$ be a trivial bundle, and
$J^\infty(\pi)$ be the bundle of its jets of the infinite order. The local coordinates on $J^\infty(\pi)$ are
$(x^i,u^\alpha,u^\alpha_I)$, where $I=(i_1, \dots, i_n)$ is a multi-index, and for every local section
$f \colon \mathbb{R}^n \rightarrow \mathbb{R}^n \times \mathbb{R}^m$ of $\pi$ the corresponding infinite jet
$j_\infty(f)$ is a section $j_\infty(f) \colon \mathbb{R}^n \rightarrow J^\infty(\pi)$ such that
$u^\alpha_I(j_\infty(f))
=\displaystyle{\frac{\partial^{\#I} f^\alpha}{\partial x^I}}
=\displaystyle{\frac{\partial^{i_1+\dots+i_n} f^\alpha}{(\partial x^1)^{i_1}\dots (\partial x^n)^{i_n}}}$.
We put $u^\alpha = u^\alpha_{(0,\dots,0)}$. Also, in the case of $m=1$ and, e.g., $n=4$ we denote
$x^1 = t$, $x^2= x$, $x^3= y$, $x^4=z$, and $u^1_{(i,j,k,l)}=u_{{t \dots t}{x \dots x}{y \dots y}{z \dots z}}$
with $i$  times $t$, $j$  times $x$, $k$  times $y$, and $l$  times $z$.

\vskip 3 pt

The vector fields
\[
D_{x^k} = \frac{\partial}{\partial x^k} + \sum \limits_{\# I \ge 0} \sum \limits_{\alpha = 1}^m
u^\alpha_{I+1_{k}}\,\frac{\partial}{\partial u^\alpha_I},
\qquad k \in \{1,\dots,n\},
\]
with $I+1_k =(i_1,\dots, i_k,\dots, i_n)+1_k = (i_1,\dots, i_k+1,\dots, i_n)$  are referred to as
{\it total derivatives}. They commute everywhere on $J^\infty(\pi)$:  $[D_{x^i}, D_{x^j}] = 0$.

\vskip 10 pt

A system of {\sc pde}s $F_r(x^i,u^\alpha_I) = 0$, $\# I \le s$, $r \in \{1,\dots, \sigma\}$, of the order
$s \ge 1$ with $\sigma \ge 1$ defines the submanifold
$\EuScript{E} = \{(x^i,u^\alpha_I) \in J^\infty(\pi) \,\,\vert\,\, D_K(F_r(x^i,u^\alpha_I)) = 0, \,\, \# K \ge 0\}$
in $J^\infty(\pi)$.

\vskip 10 pt

Denote $\mathcal{W} = \mathbb{R}^\infty$ with  coordinates $w^a$, $a \in  \mathbb{N} \cup \{0\}$. Locally,
an (infinite-di\-men\-si\-o\-nal)  {\it differential covering} over $\EuScript{E}$ is a trivial bundle
$\tau \colon J^\infty(\pi) \times \mathcal{W} \rightarrow J^\infty(\pi)$
equipped with the {\it extended total derivatives}
\begin{equation}
\tilde{D}_{x^k} = D_{x^k} + \sum \limits_{a=0}^\infty
T^a_k(x^i,u^\alpha_I,w^b)\,\frac{\partial}{\partial w^a}
\label{extended_derivatives}
\end{equation}
such that $[\tilde{D}_{x^i}, \tilde{D}_{x^j}]=0$ for all $i \not = j$ whenever
$(x^i,u^\alpha_I) \in \EuScript{E}$. For the partial derivatives of $w^a$ which are defined as
$w^a_{x^k} =  \tilde{D}_{x^k}(w^a)$ we have the system of
{\it covering equations}
\[
w^a_{x^k} = T^a_k(x^i,u^\alpha_I,w^b).
\]
This over-determined system of {\sc pde}s is compatible whenever $(x^i,u^\alpha_I) \in \EuScript{E}$.

\vskip 5 pt

Dually the covering with extended total derivatives (\ref{extended_derivatives}) is defined by the differential
ideal generated by the {\it Wahlquist--Estabrook forms}, \cite[p.~81]{LNM810},
\[
\varpi^a = dw^a - \sum \limits_{k=1}^n T^a_k(x^i,u^\alpha_I,w^b)\,dx^k.
\]
This ideal is integrable on $\EuScript{E}$, that is,
\[
d \varpi^a \equiv \sum \limits_{b} \eta^{a}_{b} \wedge \varpi^b
\,\,\,\mathrm{mod}\,\, \langle\, \vartheta_I\,\rangle,
\]
where $\eta^a_b$ are some 1-forms on $\EuScript{E}\times \mathcal{W}$ and
$\vartheta_I = (du^\alpha_I-\sum \limits_{k} u^\alpha_{I+1_k}dx^k)\vert_{\EuScript{E}}$.

\subsection{Exotic cohomology}

Let $\mathfrak{g}$ be a Lie algebra over $\mathbb{R}$ and $\rho \colon \mathfrak{g} \rightarrow \mathrm{End}(V)$
be its representation. Let $C^k(\mathfrak{g}, V) =\mathrm{Hom}(\Lambda^k(\mathfrak{g}), V)$, $k \ge 1$,
be the space of all $k$--linear skew-symmetric mappings from $\mathfrak{g}$ to $V$. Then
the Chevalley--Eilenberg differential
complex
\[
V=C^0(\mathfrak{g}, V) \stackrel{d}{\longrightarrow} C^1(\mathfrak{g}, V)
\stackrel{d}{\longrightarrow} \dots \stackrel{d}{\longrightarrow}
C^k(\mathfrak{g}, V) \stackrel{d}{\longrightarrow} C^{k+1}(\mathfrak{g}, V)
\stackrel{d}{\longrightarrow} \dots
\]
is is generated by the differential defined by the formula
\[
%\fl
d \theta (X_1, ... , X_{k+1}) =
\sum\limits_{q=1}^{k+1}
(-1)^{q+1} \rho (X_q)\,(\theta (X_1, ... ,\hat{X}_q, ... ,  X_{k+1}))
\]
\begin{equation}
\quad
+\sum\limits_{1\le p < q \le k+1} (-1)^{p+q}
\theta ([X_p,X_q],X_1, ... ,\hat{X}_p, ... ,\hat{X}_q, ... ,  X_{k+1}).
\label{CE_differential}
\end{equation}
The cohomology groups of the complex $(C^{*}(\mathfrak{g}, V), d)$ are referred to as
the {\it cohomology groups of the Lie algebra} $\mathfrak{g}$ {\it with coefficents in the representation}
$\rho$. For the trivial representation $\rho_0 \colon \mathfrak{g} \rightarrow \mathbb{R}$,
$\rho_0 \colon X \mapsto 0$, the complex and its cohomology are denoted by $C^{*}(\mathfrak{g})$ and
$H^{*}(\mathfrak{g})$, respectively.

Consider a Lie algebra $\mathfrak{g}$ over $\mathbb{R}$ with non-trivial first cohomology group
$H^1(\mathfrak{g})$ and take a closed 1-form $\alpha$ on $\mathfrak{g}$. Then for any $\lambda \in \mathbb{R}$
define new differential
$d_{\lambda\alpha} \colon C^k(\mathfrak{g},\mathbb{R}) \rightarrow C^{k+1}(\mathfrak{g},\mathbb{R})$ by
the formula
\[
d_{\lambda \alpha} \theta = d \theta +\lambda \,\alpha \wedge \theta.
\]
From  $d\alpha = 0$ it follows that
\begin{equation}
d_{\lambda \alpha} ^2=0.
\label{d_deformed_2}
\end{equation}
The cohomology groups of the complex
\[
C^1(\mathfrak{g}, \mathbb{R})
\stackrel{d_{\lambda \alpha}}{\longrightarrow}
\dots
\stackrel{d_{\lambda \alpha}}{\longrightarrow}
C^k(\mathfrak{g}, \mathbb{R})
\stackrel{d_{\lambda \alpha}}{\longrightarrow}
C^{k+1}(\mathfrak{g}, \mathbb{R})
\stackrel{d_{\lambda \alpha}}{\longrightarrow} \dots
\]
are referred to as the {\it exotic} {\it cohomology groups} of $\mathfrak{g}$ and denoted by
$H^{*}_{\lambda\alpha}(\mathfrak{g})$.

\vskip 5 pt

\noindent
{\sc Remark 1}.
Cohomology $H^{*}_{\lambda\alpha}(\mathfrak{g})$ coincides with cohomology of $\mathfrak{g}$ with coefficients
in the one-dimensional representation $\rho_{\lambda\alpha} \colon \mathfrak{g} \rightarrow \mathbb{R}$,
$\rho_{\lambda\alpha} \colon X \mapsto \lambda\, \alpha(X)$. In particular,
when $\lambda=0$, cohomology $H^{*}_{\lambda\alpha}(\mathfrak{g})$ coincides with $H^{*}(\mathfrak{g})$.
\hfill$\diamond$

\vskip 5 pt

\noindent
{\sc Remark 2}.
In all the cases considered in this paper
$H^1(\mathfrak{g})=Z^1(\mathfrak{g})$ due to
$C^0(\mathfrak{g}) = \{0\}$ and $B^1(\mathfrak{g})=\{0\}$,
so a closed 1-form $\alpha$ can be identified with its cohomology class.
\hfill$\diamond$

%%%%%%%%%%%%%%%%%%%%%%%%%%%%%%%%%%%%%%%%%%%%%%%%%%%%%%%%%%%%%%%%%%%%%%%%%%%%%%%%%%%%%%%%%%%%%%%%%%

\section{The Lax representation for the potential Khokhlov-Zabolotskaya equation
through the second exotic cohomology of the symmetry pseudogroup}
\label{pKhZ_section}

A relation between the exotic cohomology of symmetry pseudogroups and Lax re\-pre\-sen\-ta\-ti\-ons for
integrable systems was established in \cite{Morozov2017}. In a slightly different notation one of the results
of that paper can be presented as follows.

Consider the potential Khokhlov--Za\-bo\-lot\-ska\-ya equation $\EuScript{E}_1$
(or Lin--Reis\-sner--Tsien equation), \cite{LinReissnerTsien,KhZ},
\begin{equation}
u_{yy} = u_{tx} + u_x\,u_{xx}.
\label{pKhZ}
\end{equation}
The infinite normal prolongation,
\cite{Cartan1,Cartan4,Vasilieva1972,Stormark2000},  of the
structure equations for the pseudogroup $Sym(\EuScript{E}_1)$ of  contact symmetries of this equation
has the form
\begin{eqnarray}
\fl
d\alpha &=& 0,
\label{sym_pKhZ_d_alpha}
\\
\fl
d\Theta_0 &=&
\hspace{57pt}
\nabla \Theta_0 \wedge \Theta_0,
\label{sym_pKhZ_0_series}
%\phantom{\frac{\frac{A}{A}}{\frac{A}{A}}}
\\
\fl
d\Theta_1
&=& \,\,\,\,
\alpha \wedge \Theta_1
+
\nabla \Theta_1 \wedge \Theta_0
+
\frac{2}{3}\,\nabla \Theta_0 \wedge \Theta_1,
%\phantom{\frac{\frac{A}{A}}{\frac{A}{A}}}
\label{sym_pKhZ_1_series}
\\
\fl
d\Theta_2 &=& 2\,\alpha \wedge \Theta_2 +\nabla \Theta_2 \wedge \Theta_0
+\frac{2}{3}\,\nabla \Theta_1 \wedge \Theta_1+\frac{1}{3}\,\nabla \Theta_0 \wedge \Theta_2,
%\phantom{\frac{\frac{A}{A}}{\frac{A}{A}}}
\label{sym_pKhZ_2_series}
\\
\fl
d\Theta_3 &=& 3\,\alpha \wedge \Theta_3 + \nabla \Theta_3 \wedge \Theta_0
+\frac{2}{3}\,\nabla \Theta_2 \wedge \Theta_1+\frac{1}{3}\,\nabla \Theta_1 \wedge \Theta_2,
%\phantom{\frac{\frac{A}{A}}{\frac{A}{A}}}
\label{sym_pKhZ_3_series}
\\
\fl
d\Theta_4 &=& 4\,\alpha \wedge \Theta_4 +\, \nabla \Theta_4 \wedge \Theta_0
+\frac{2}{3}\,\nabla \Theta_3 \wedge \Theta_1 +\frac{1}{3}\,\nabla \Theta_2 \wedge \Theta_2
%--------------------------------------------------------------------------------------------------
\nonumber
\\
&&
%--------------------------------------------------------------------------------------------------
-\frac{1}{3}\,\nabla \Theta_0 \wedge \Theta_4,
\label{sym_pKhZ_4_series}
%\phantom{\frac{\frac{A}{A}}{\frac{A}{A}}}
\end{eqnarray}
where
\begin{equation}
\Theta_m = \sum \limits_{j=0}^{\infty}
\displaystyle{\frac{h^j}{j !}} \,\theta_{m,j},
\qquad
\nabla \Theta_m = \displaystyle{\frac{\partial}{\partial h}} \Theta_m =
\sum \limits_{j=0}^{\infty} \displaystyle{\frac{h^j}{j !}} \,\theta_{m,j+1}
\label{big_theta}
\end{equation}
for $0 \le m \le 4$,
while $dh =0$ and $\theta_{3,0}=0$. We have
\begin{eqnarray*}
\alpha &=& p^{-1}\, dp,
\qquad
\theta_{0,0} = q\,dt,
\\
\theta_{1,0} &=& p\,q^{2/3} \left(dy + a_1\,dt\right),
\\
\theta_{2,0} &=& p^2\,q^{1/3} \left(dx + \textfrac{2}{3}\,a_1\,dy+a_2\,dt\right),
\\
\theta_{4,0} &=& p^4\,q^{-1/3} \left(du-u_t\,dt-u_x\,dx- u_y\,dy\right),
\end{eqnarray*}
where $p, q, a_1, a_2 \in \mathbb{R}$, $p\neq 0$,  $q\neq 0$, are parameters.

\vskip 5 pt
\noindent
{\sc Remark} 3.
All the other forms $\theta_{i,j}$ can be found inductively from the series of equations
(\ref{sym_pKhZ_d_alpha}) -- (\ref{sym_pKhZ_4_series}). For example, the first equation from the series of
equations  (\ref{sym_pKhZ_0_series}), $d\theta_{0,0} = \theta_{0,1} \wedge \theta_{0,0}$, has the form
$dq \wedge dt = \theta_{0,1} \wedge q\,dt$, therefore we get
$\theta_{0,1} = q^{-1}\,dq +b_1\,dt$ with $b_1 \in \mathbb{R}$.
Now forms $\theta_{0,0}$ and $\theta_{0,1}$ in the second equation
$d\theta_{1,0} = \theta_{0,2} \wedge \theta_{0,0}$ from the series (\ref{sym_pKhZ_0_series}) are known, and we
obtain $\theta_{0,2} = db_1 +b_2\,dt$, $b_2 \in \mathbb{R}$. Likewise, we can compute all the forms
$\theta_{0,j}$. When  $\theta_{0,j}$ and $\theta_{1,0}$ are known, from (\ref{sym_pKhZ_1_series})
it is possible to find $\theta_{1,j}$, etc.
\hfill $\diamond$
\vskip 5 pt
Equations (\ref{sym_pKhZ_d_alpha}) -- (\ref{sym_pKhZ_4_series}) imply
$H^1(Sym(\EuScript{E}_1)) = \mathbb{R} \,\alpha$. The following theorem describes the structure of the second
exotic cohomology group for $Sym(\EuScript{E}_1)$, for the proof of a more general result see Theorem 2 below.
\vskip 5pt
\noindent
{\sc Theorem} 1.
\[
H^2_{\lambda\,\alpha} (Sym(\EuScript{E}_1)) =
\left\{
\begin{array}{lcl}
\mathbb{R}\,[\Omega],& ~~~& \lambda = -3,
\\
\{0\},& ~~~& \lambda \neq -3,
\end{array}
\right.
\]
where
$\Omega = \theta_{3,1} \wedge \theta_{0,0} + \textfrac{2}{3}\,\theta_{2,1} \wedge \theta_{1,0}
+\textfrac{1}{3}\,\theta_{1,1} \wedge \theta_{2,0}$.
\vskip 10 pt
\noindent
{\sc Corollary}. Equation
\begin{equation}
d\omega = 3\,\alpha \wedge \omega +\Omega
\label{omega_pKhZ}
\end{equation}
is compatible with the structure equations
(\ref{sym_pKhZ_d_alpha}) -- (\ref{sym_pKhZ_4_series})
of $Sym(\EuScript{E}_1)$.
\vskip 10 pt
\noindent
{\sc Remark} 4.
If we rename $\omega=\theta_{3,0}$, then (\ref{omega_pKhZ}) gets the form
\begin{equation}
d\theta_{3,0} = 3\,\alpha \wedge \theta_{3,0}
+ \theta_{3,1} \wedge \theta_{0,0}
+\frac{2}{3}\,\theta_{2,1} \wedge \theta_{1,0}+\frac{1}{3}\,\theta_{1,1} \wedge \theta_{2,0}.
\label{d_theta_30_pKhZ}
\end{equation}
Therefore, if we add $\theta_{3,0}$ as the coefficient at $h^0$ in the series for
$\Theta_3$ in (\ref{big_theta}), then  (\ref{sym_pKhZ_3_series}) remains valid.
\hfill$\diamond$

\vskip 5 pt
Since equation (\ref{d_theta_30_pKhZ}) is compatible with system  (\ref{sym_pKhZ_d_alpha})--(\ref{sym_pKhZ_4_series}),
Lie's third inverse fundamental theorem in Cartan's form, \cite{Cartan1,Cartan4,Vasilieva1972,Stormark2000},
ensures existence of a solution $\theta_{3,0}$ to (\ref{d_theta_30_pKhZ}).
We integrate this equation and put $a_1=v_x$.  This yields
\[
\theta_{3,0} = p^3\,\left(dv-v_x\,dx - \left(\textfrac{1}{3}\,v_x^3 - u_x\,v_x - u_y\right)\,dt
-\left(\textfrac{1}{2}\,v_x^2 - u_x\right)\,dy
\right).
\]
This is the Wahlquist--Estabrook form of the Lax representation,
\cite{Kuzmina1967,Gibbons1984,Krichever1990,Zakharov1994},
\begin{equation}
\left\{
\begin{array}{lcl}
v_t &=& \textfrac{1}{3}\,v_x^3 - u_x\,v_x - u_y,
\phantom{\frac{A}{A}}
\\
v_y &=& \textfrac{1}{2}\,v_x^2 - u_x
%\phantom{\frac{A}{A}}
\phantom{\frac{\frac{A}{A}}{\frac{A}{A}}}
\end{array}
\right.
\label{KhZ_covering}
\end{equation}
for equation (\ref{pKhZ}).

%%%%%%%%%%%%%%%%%%%%%%%%%%%%%%%%%%%%%%%%%%%%%%%%%%%%%%%%%%%%%%%%%%%%%%%%%%%%%%%%%%%%%%%%%%%%%%%%%%%%%%

\section{A deformation of the tensor product of the Lie algebra of vector fields on a line and the algebra of
truncated polynomials}
\label{section4}

The structure equations (\ref{sym_pKhZ_d_alpha})--(\ref{sym_pKhZ_4_series}), (\ref{d_theta_30_pKhZ})
can be written in the form
\begin{eqnarray*}
d\alpha&=&0,
\\
d\Theta_k &=& k\,\alpha \wedge \Theta_k + \sum \limits_{m=0}^k \left(1-\textfrac{1}{3}\,m\right)
\,\nabla\Theta_{k-m}\wedge \Theta_m,
\qquad 0\le k \le 4.
\end{eqnarray*}
Then the results of Section \ref{pKhZ_section} admit the following generalization.
\vskip 5 pt
\noindent
{\sc Definition}. For $n \in \mathbb{N}$ and $\varepsilon \in \mathbb{R}$
denote by
$\mathfrak{G}(n,\varepsilon)$
the Lie algebra with the structure equations
\begin{eqnarray}
d\alpha&=&0,
\nonumber %\label{d_alpha}
\\
d\Theta_k &=& k\,\alpha \wedge \Theta_k + \sum \limits_{m=0}^k \left(1+\varepsilon\,m\right)
\,\nabla \Theta_{k-m}\wedge \Theta_m,
\quad
0 \le  k \le n,
\label{structure_eqns_of_G_n_epsilon}
\end{eqnarray}
where equations (\ref{big_theta})
hold for each $m\in \{0,1,\dots,n\}$.
\vskip 5 pt

\noindent
{\sc Remark} 5.
The Lie algebra $\mathfrak{G}(n, \varepsilon)$ has the following description. Consider the tensor product
$\mathbb{R}_{n+1}[s] \otimes C^\omega(\mathbb{R})$ of the algebra
$\mathbb{R}_{n+1}[s] = \mathbb{R}[s] / \langle s^{n+1} \rangle$ of truncated polynomials of degree less that
$n+1$ and the algebra of real-analytic functions of $t \in \mathbb{R}$. This is a vector space over $\mathbb{R}$
generated by functions $X_{k,m} = \frac{1}{m!}\,s^k\,t^m$, $k \in \{0, \dots, n\}$, $m \in \mathbb{N} \cup \{0\}$.
This vector space is equipped with the Lie bracket
\begin{equation}
[f,g]_{\varepsilon} = f g_t-g  f_t+\varepsilon\,s\,(f_s g_t-g_s f_t).
\label{epsilon_bracket}
\end{equation}
The vector field $Y = s\,\partial_s$  is an outer derivative of the above Lie algebra. Then
$\mathfrak{G}(n, \varepsilon)$ is the semi-direct sum of
$\mathbb{R}_{n+1}[s] \otimes C^\omega(\mathbb{R})$ and one-dimensional Lie algebra generated by $Y$:
\[
\mathfrak{G}(n, \varepsilon)= \left(\mathbb{R}_{n+1}[s] \otimes C^\omega(\mathbb{R})\right) \rtimes \mathbb{R}\, Y.
\]
This Lie algebra is a deformation of $\mathfrak{G}(n,0)$. For $\varepsilon = 0$ the bracket
(\ref{epsilon_bracket}) is the Lie bracket in the algebra of analytic vector fields on a line.

Consider 1-forms $\theta_{i,j}$, $\alpha$
dual to $X_{i,j}$, $Y$, that is, 1-forms on $\mathfrak{G}(n,\varepsilon)$ such that equations
\begin{equation}
\theta_{i,j}(X_{k,m}) = \delta_{ik}\,\delta_{jm},
\quad
\theta_{i,j}(Y)=0,
\quad
\alpha(X_{k,m})=0,
\quad
\alpha(Y)=1
\label{dual_forms}
\end{equation}
hold for all $i,j, k, m \in \mathbb{N} \cup \{0\}$. Then  (\ref{dual_forms}), (\ref{CE_differential}), and
(\ref{big_theta}) imply the structure equ\-a\-ti\-ons
(\ref{structure_eqns_of_G_n_epsilon}).
\hfill$\diamond$
\vskip 7 pt

System (\ref{structure_eqns_of_G_n_epsilon}) is the infinite normal prolongation of the system
\begin{eqnarray}
d\alpha&=&0,
\label{d_alpha}
\\
d\theta_{k,0} &=& k\,\alpha \wedge \theta_{k,0}
+ \sum \limits_{m=0}^k \left(1+\varepsilon\,m\right)
\,\theta_{k-m,1}\wedge \theta_{m,0},
\quad
0 \le  k \le n.
\label{structure_eqns_of_G_n_epsilon_pseudogroup}
\end{eqnarray}
This system is involutive, \cite[\S 6]{Cartan1}, \cite[Def. 11.7]{Olver1995}, therefore the third inverse
fundamental Lie's theorem in Cartan's form implies existence of 1-forms $\alpha$, $\theta_{k,j}$,
$k \in \{0, \dots, n\}$, $j \in \{0,1\}$, that satisfy  (\ref{structure_eqns_of_G_n_epsilon_pseudogroup}).
Forms $\alpha$, $\theta_{0,0}$, ... , $\theta_{n,0}$ define a Lie pseudo-group on $\mathbb{R}^{n+1}$. While for
the purposes of the present paper we need explicit expressions for the forms $\theta_{k,0}$ and $\theta_{k,1}$
only, all the other forms $\theta_{k,j}$ can be found inductively by integration of equations
(\ref{structure_eqns_of_G_n_epsilon}). We need the whole system (\ref{structure_eqns_of_G_n_epsilon}) to prove
Theorem 2 below.

We note that system (\ref{structure_eqns_of_G_n_epsilon_pseudogroup}) has the following specific structure.
For each fixed $k \in \{0, \dots, n\}$ the first $k+1$ equations from
(\ref{structure_eqns_of_G_n_epsilon_pseudogroup}) satisfy the conditions of the Frobenius theorem, see  e.g.
\cite[Th.~1.3.4]{IveyLandsberg},  for forms $\theta_{0,0}$, ..., $\theta_{k,0}$. We consider $\alpha$ and
$\theta_{j,0}$, $\theta_{j,1}$ for $j \in \{0, \dots, k\}$  as 1-forms on $\mathbb{R}^N$ for $N \ge 3\,(n+1)+1$.
Eq. (\ref{d_alpha}) implies that there is a function $p\neq 0$ on $\mathbb{R}$ such
that\footnote{Here and below we put $\alpha =dp/p$ instead of the natural choice $\alpha = dp$ to simplify
the further computations} $\alpha= dp/p$. Then the Frobenius theorem implies existence of independent functions
(coordinates) $x_0$, ... , $x_{k}$ on $\mathbb{R}^N$ such that the differential ideal of forms $\theta_{0,0}$,
... , $\theta_{k,0}$ is generated algebraically by forms $dx_0$, ... , $dx_k$, that is,
$\theta_{0,0} = A_0\,dx_0$ and
$\theta_{i,0} = A_i\,\left (dx_i+\sum\limits_{j=0}^{i-1} B_{ij} dx_j\right)$, $i \in \{1,\dots, k\}$,
for some functions $A_j \ne 0$, $B_{ij}$ on $\mathbb{R}^N$.
Substituting these forms into (\ref{structure_eqns_of_G_n_epsilon_pseudogroup}) yields a triangular
system for 1-forms $\theta_{i,1}$, so it is easy to find them.
\vskip 5 pt
\noindent
\addtocounter{example_counter}{1}
{\sc Example}
\arabic{example_counter}.
For $k=2$ from first three equations of (\ref{structure_eqns_of_G_n_epsilon_pseudogroup}) we have
\begin{eqnarray*}
\theta_{0,0} &=& a_0 \,dx_0,
\\
\theta_{0,1} &=& p\,a_0^{1+\varepsilon} \,(dx_1+a_1\,dx_0),
\\
\theta_{0,2} &=& p^2\,a_0^{1+2\varepsilon}\,(dx_2+(1+\varepsilon)\,a_1\,dx_1+a_2\,dx_0),
\\
\theta_{0,1} &=& da_0/a_0+b_0\,dx_0,
\\
\theta_{1,1} &=& p\,a_0^{\varepsilon}\,(da_1 +(1+\varepsilon)\,b_0\,dx_1+b_1\,dx_0),
\\
\theta_{2,1}  &=& p^2\,a_0^{2\varepsilon}\,(da_2-(1+\varepsilon)\,a_1 da_1 +(1+2\,\varepsilon)\,b_0 dx_2
\\
&&+(1+\varepsilon)\,(b_1+\varepsilon\,a_1 b_0)\,dx_1+b_2\,dx_2)
\end{eqnarray*}
with $a_0 \neq 0$.
Then substituting for $\theta_{3,0} = A_3\,(dx_3+B_{30}dx_0+B_{31} dx_1+B_{32} dx_2)$ into
the equation for $d\theta_{3,0}$ from (\ref{structure_eqns_of_G_n_epsilon_pseudogroup})
allows one to define coefficients $A_3$, $B_{3j}$. After this, the equation for $d\theta_{3,0}$
defines $\theta_{3,1}$ up to adding the form $b_3\,dx_0$ with the new coordinate $b_3$ on $\mathbb{R}^N$.
\hfill$\diamond$

\vskip 7 pt

The Lie algebra $\mathfrak{G}(n, \varepsilon)$ has the following important feature: when
$\varepsilon =-\sfrac{1}{r}$ for $r \in \mathbb{N}$ and $n  \ge r$, 1-form $\theta_{r,0}$ has only two
entries in the whole system (\ref{structure_eqns_of_G_n_epsilon}). Indeed, the first equation from the series
for $d\Theta_r$ acquires the form
\begin{equation}
d\theta_{r,0} = r\,\alpha \wedge \theta_{r,0}
+\Psi_r,
\label{d_theta_r_0}
\end{equation}
while neither 2-form $\Psi_r$ nor the other
equations from  (\ref{structure_eqns_of_G_n_epsilon}) include $\theta_{r,0}$.
Therefore in order to generalize results of Section \ref{pKhZ_section} we consider the Lie algebra
$\mathfrak{G}(r+1,\sfrac{-1}{r})$ and proceed in the following steps. First,
we find forms $\theta_{i,0}$ and $\theta_{i,1}$ for  $0 \le i \le r-1$ from
equations (\ref{structure_eqns_of_G_n_epsilon}) in the similar way as it was described above.
Second,  we assume that $\theta_{r+1,0} = A_{r+1}\,(du-u_{x_0}\, dx_0-\dots -u_{x_{r-1}}\,dx_{r-1})$,
$A_{r+1}\neq 0$, that is, we consider $\theta_{r+1,0}$ as the zeroth order contact form on the jet bundle
$J^{\infty}(\pi)$ for the bundle
$\pi \colon (x_0, x_1, \dots, x_{r-1}, u) \mapsto (x_0, x_1, \dots, x_{r-1})$.
This implies some expressions for the parameters $B_{i,j}$  in terms of
$u_{x_0}$, $\dots$, $u_{x_{r-1}}$. Third, we integrate equation (\ref{d_theta_r_0}) to find
$\theta_{r,0}= b_{r}\,(dv-T_0\, dx_0-\dots -T_{r-1}\,dx_{r-1})$
and then consider the system of {\sc pde}s
$v_{x_i} = T_i$, $0 \le i \le r-1$, generated by equation $\theta_{r,0}=0$.

%%%%%%%%%%%%%%%%%%%%%%%%%%%%%%%%%%%%%%%%%%%%%%%%%%%%%%%%%%%%%%%%%%%%%%%%%%%%%%%%%%%%%%%%%%%%%%%%%%%%%%

\section{Lie algebras $\mathfrak{G}(r+1,\sfrac{-1}{r})$, their extensions, and associated
integrable systems}
\label{section5}

\subsection{Integrable systems generated by $\mathfrak{G}(r+1,\sfrac{-1}{r})$}

In this subsection we present some results of computations discussed in the end of Section \ref{section4}.

\vskip 5pt

\noindent
\addtocounter{example_counter}{1}
{\sc Example}
\arabic{example_counter}.
For
$\mathfrak{G}(5, \sfrac{-1}{4})$
we have\footnote{Here and below we rescale parameters in forms $\theta_{j,0}$ in order to simplify the resulting
{\sc pde} and its Lax representation.}
\[
\theta_{0,0} = \textfrac{1}{4}\,q^4\,dt,
\]
\[
\theta_{1,0} = \textfrac{1}{3}\,p\,q^3\,(dy + a\,dt),
\]
\[
\theta_{2,0} = \textfrac{1}{2}\,p^2\,q^2\,(dx + 2\,a\,dy + (3\,a^2-2\,u_z)\,dt),
\]
\[
\theta_{3,0} = p^3\,q\,(dz + a\,dx + (a^2-u_z)\,dy + (a^3-2\,a\,u_z-u_x)\,dt),
\]
\[
\theta_{5,0} = p^5\,q^{-1}\,(du -u_t\,dt -u_x\,dx-u_y\,dy-u_z\,dz),
\]
Then we integrate (\ref{d_theta_r_0}) with $r=4$ and put $a = v_z$. This gives
\[
\theta_{4,0} = -p^4\,\left(
dv
-\left(\textfrac{1}{4}\,v_z^4-u_z\,v_z^2-u_x\,v_z-u_y+\textfrac{1}{2}\,u_z^2\right)\,dt
-\left(\textfrac{1}{2}\,v_z^2-u_z\right)\,dx
\right.
\]
\[
\qquad\qquad
\left.
-\left(\textfrac{1}{3}\,v_z^3-u_z\,v_z-u_x\right)\,dy -v_z\,dz
\right)
\]
Then equation $\theta_{4,0}=0$ yields the system
\[
\left\{
\begin{array}{lcl}
v_x &=&  \textfrac{1}{2}\,v_z^2-u_z,
\phantom{\frac{\frac{A}{A}}{\frac{A}{A}}}
\\
v_y&=& \textfrac{1}{3}\,v_z^3-u_z\,v_z-u_x,
\phantom{\frac{\frac{A}{A}}{\frac{A}{A}}}
\\
v_t &=& \textfrac{1}{4}\,v_z^4-u_z\,v_z^2-u_x\,v_z-u_y+\textfrac{1}{2}\,u_z^2.
\phantom{\frac{\frac{A}{A}}{\frac{A}{A}}}
\end{array}
\right.
\]
This system is compatible whenever equations
\begin{equation}
\left\{
\begin{array}{lcl}
u_{xx} &=& u_{yz}+u_z\,u_{zz},
\\
u_{xy} &=& u_{tz} +u_z\,u_{xz} + u_x\,u_{zz},
\\
u_{yy} &=& u_{tx}+u_x\,u_{xz}+u_z^2\,u_{zz}
\end{array}
\right.
\label{dKP_2_system}
\end{equation}
hold. System (\ref{dKP_2_system}) is the second system from the dKP hierarchy,
\cite{Zakharov1980,Kodama1988,DubrovinNovikov1989,Krichever1990,Kupershmidt1990,Zakharov1994,
CarollKodama1995,GibbonsTsarev1996}.
\hfill $\diamond$
\vskip 5 pt

\noindent
\addtocounter{example_counter}{1}
{\sc Example}
\arabic{example_counter}.
The Lie algebra
$\mathfrak{G}(6, \sfrac{-1}{5})$
provides the third system from the dKP hierarchy. In this case integration of (\ref{structure_eqns_of_G_n_epsilon}) gives
\[
\theta_{0,0} = \textfrac{1}{5}\,q^5\,dt,
\]
\[
\theta_{1,0} = p\,q^4\,\left(\textfrac{1}{4}\,dy + a\,dt\right),
\]
\[
\theta_{2,0} = p^2\,q^3\,\left(\textfrac{1}{3}\,dx + a\,dy + (2\,a^2-u_s)\,dt\right),
\]
\[
\theta_{3,0} = p^3\,q^2\,\left(\textfrac{1}{2}\,dz + a\,dx + (3\,a^2-2\,u_s)\,dy + 2\,(2\,a^3-3\,a\,u_s-u_z)\,dt\right),
\]
\[
\theta_{4,0} = p^4\,q\,\left(ds + a\,dz+ (a^2-u_s)\,dx
+ (a^3-2\,u_s\,a-u_z)\,dy
\right.
\]
\[
\qquad\qquad
\left.
+ (a^4-3\,u_s\,a^2-2\,u_z\,a-u_x+u_s^2)\,dt\right),
\]
\[
\theta_{6,0} = p^6\,q^{-1}\,(du -u_t\,dt -u_x\,dx-u_y\,dy-u_z\,dz-u_s\,ds).
\]
Then from (\ref{d_theta_r_0}) we get
\[
\theta_{5,0} = -p^5\,\left(
dv
-\left(
\textfrac{1}{5}\,v_s^5-u_s\,v_s^3-u_z\,v_s^2+(u_s^2-u_x)\,v_s+u_z\,u_s-u_y
\right)\,dt
\right.
\]
\[
\qquad\qquad
-\left(\textfrac{1}{3}\,v_s^3-u_s\,v_s-u_z\right)\,dx
-\left(\textfrac{1}{4}\,v_s^4-u_s\,v_s^2-u_z\,v_s-u_x+\textfrac{1}{2}\,u_s^2\right)\,dy
\]
\[
\qquad\qquad
\left.
 -\left(\textfrac{1}{2}\,v_s^2-u_s\right)\,dz
-v_s\,ds
\right)
\]
with $a=v_s$.  This form generates the system
\[
\left\{
\begin{array}{lcl}
v_z&=& \textfrac{1}{2}\,v_s^2-u_s,
\phantom{\frac{\frac{A}{A}}{\frac{A}{A}}}
\\
v_x &=& \textfrac{1}{3}\,v_s^3-u_s\,v_s-u_z,
\phantom{\frac{\frac{A}{A}}{\frac{A}{A}}}
\\
v_y &=&  \textfrac{1}{4}\,v_s^4-u_s\,v_s^2-u_z\,v_s-u_x+\textfrac{1}{2}\,u_s^2,
\phantom{\frac{\frac{A}{A}}{\frac{A}{A}}}
\\
v_t &=& \textfrac{1}{5}\,v_s^5-u_s\,v_s^3-u_z\,v_s^2+(u_s^2-u_x)\,v_s+u_z\,u_s-u_y.
\phantom{\frac{\frac{A}{A}}{\frac{A}{A}}}
\end{array}
\right.
\]
This is the Lax representation for the third system from the dKP hierarchy
\[
\left\{
\begin{array}{lcl}
u_{zz} &=& u_{xs}+u_s\,u_{ss},
\\
u_{xz} &=& u_{ys}+u_s\,u_{zs}+u_z\,u_{ss},
\\
u_{yz} &=& u_{ts}+u_s\,u_{xs}+u_z\,u_{zs}+u_x\,u_{ss},
\\
u_{xx} &=& u_{yz}+u_z\,u_{zs}+u_s^2\,u_{ss},
\\
u_{xy} &=& u_{tz}+u_z\,u_{xs}+(u_x+u_s^2)\,u_{zs}+2\,u_z\,u_s\,u_{ss},
\\
u_{yy} &=& u_{tx}+u_x\,u_{xs}+2\,u_z\,u_s\,u_{zs}+(u_z^2+u_s^3)\,u_{ss}.
\end{array}
\right.
\]
\hfill$\diamond$

\subsection{Right extensions of $\mathfrak{G}(n,\sfrac{-1}{r})$}

While it is natural to expect that Maurer-Cartan forms of $\mathfrak{G}(r+1,\sfrac{-1}{r})$ with $r>5$
produce higher elements of the dKP hierarchy, an interesting question is to consider the case $r=2$. But
the integration scheme from the end of Section \ref{section4}  does not give any Lax representation of a
{\sc pde}
in the case of
 %for
$\mathfrak{G}(3,\sfrac{-1}{2})$. Therefore we consider an extension   of
$\mathfrak{G}(n,\varepsilon)$. For $n \ge r \ge 1$ we denote by
$\mathfrak{H}(n,r)$ the Lie algebra with the structure e\-qu\-a\-ti\-ons
\begin{eqnarray}
\fl
d\alpha&=&0,
\nonumber
\\
\fl
d\beta&=&r\,\alpha \wedge \beta,
\nonumber
\\
\fl
d\Theta_{k^\prime} &=& {k^\prime}\,\alpha \wedge \Theta_{k^\prime}
+ \sum \limits_{m=0}^{k^\prime} \left(1-\frac{m}{r}\right)
\,\nabla \Theta_{k^\prime -m}\wedge \Theta_m,
\nonumber
\\
\fl
d\Theta_{k^{\prime\prime}} &=& {k^{\prime\prime}}\,\alpha \wedge \Theta_{k^{\prime\prime}}
+ \sum \limits_{m=0}^{k^{\prime\prime}} \left(1-\frac{m}{r}\right)
\,\nabla \Theta_{k^{\prime\prime} -m}\wedge \Theta_m +\beta \wedge \nabla \Theta_{k^{\prime\prime}-r},
\label{structure_eqns_of_H_n_epsilon_r}
\end{eqnarray}
where $k^\prime \in \{0, \dots, r-1\}$ and $k^{\prime\prime} \in\{r, \dots,  n\}$.
Then $\mathfrak{H}(n,r)$ is a one-dimensional right extension, \cite[\S~1.4.4]{Fuks1984},
of $\mathfrak{G}(n,\sfrac{-1}{r})$. Indeed, from the structure equations
(\ref{structure_eqns_of_H_n_epsilon_r}) it follows that the dual element associated to the
new 1-form $\beta$ is a derivative
$Z \colon \mathfrak{G}(n,\sfrac{-1}{r}) \rightarrow \mathfrak{G}(n,\sfrac{-1}{r})$ such that
$Z \colon X_{k,m} \mapsto X_{k+r,m-1}$ for $m \ge 1$ and $k+r \le n$,
$Z \colon X_{k,m} \mapsto 0$ for $m=0$ or $k+r > n$,
and $Z(Y(f))-Y(Z(f))=-r\,Y(f)$ for every $f \in \mathfrak{G}(n,\sfrac{-1}{r})$,
while there is no function $g \in \mathbb{R}_{n+1}[s] \otimes C^\omega(\mathbb{R})$
such that $[g, X_{k,m}]_{-1/r}=X_{k+r,m-1}$ for all $m \ge 1$ and $k+r \le n$.

\vskip 10 pt

\noindent
\noindent
\addtocounter{example_counter}{1}
{\sc Example}
\arabic{example_counter}.
Consider the Lie algebra $\mathfrak{H}(3,2)$. We have
\[
H^2_{\lambda\,\alpha} (\mathfrak{H}(3,2)) =
\left\{
\begin{array}{lcl}
\mathbb{R}\,[\alpha \wedge \beta], & ~~~~& \lambda = -2,
\\
\{0\}, & ~~~~& \lambda \neq -2.
\end{array}
\right.
\]
This assertion can be proven similarly to Theorem 2 below.
Integration  yields
\[
\beta = p^2\,dx,
\]
\[
\theta_{0,0} = -\textfrac{1}{2}\,q^2\,dt,
\]
\[
\theta_{1,0} = p\,q\,\left(dy -u_x\,dt\right),
\]
\[
\theta_{3,0} = p^3\,q^{-1}\,(du -u_t\,dt -u_x\,dx-u_y\,dy).
\]
Then instead of equation
$d\theta_{2,0} = 2\,\alpha \wedge \theta_{2,0} +\Phi$
with $\Phi=
\theta_{2,1}\wedge \theta_{0,0}
+\textfrac{1}{2}\,\theta_{1,1}\wedge \theta_{1,0}
+\beta\wedge \theta_{0,1}
$
from (\ref{structure_eqns_of_H_n_epsilon_r}) we consider equation
\[
d\tilde{\theta}_{2,0} = 2\,\alpha \wedge \tilde{\theta}_{2,0}
+\Phi+2\,\alpha \wedge \beta.
\]
We rename $q=p\,\mathrm{exp}(v_x)$ and get
\[
\tilde{\theta}_{2,0} = p^2\,\left(dv+\left(u_y-\textfrac{1}{2}\,u_x^2\right)\,dt-v_x\,dx+u_x\,dy
\right).
\]
This form generates the system
\begin{equation}
\left\{
\begin{array}{lcl}
v_t &=& \textfrac{1}{2}\,u_x^2-u_y,
\\
v_y &=& -u_x.
\end{array}
\right.
\label{conservation_law_for_Gerdjikov_eq}
\end{equation}
This system is compatible whenever the equation \cite{Gerdjikov1990,Blaszak2002,Pavlov2003}
\begin{equation}
u_{yy} = u_{tx}+u_x\,u_{xy}.
\label{Gerdjikov_eq}
\end{equation}
holds.  Thus system (\ref{conservation_law_for_Gerdjikov_eq}) defines an Abelian covering,
\cite{VinogradovKrasilshchik1997}, or a conservation law, for equation (\ref{Gerdjikov_eq}).
To obtain non-Abelian covering for (\ref{Gerdjikov_eq}) we consider the form
\begin{eqnarray}
\fl
\omega&=&
\tilde{\theta}_{2,0}+\theta_{0,0}-\theta_{1,0} =
p^2\,\left(dv
+\left(u_y-\textfrac{1}{2}\,u_x^2-\textfrac{1}{2}\,\mathrm{e}^{2\,v_x}+u_x\,\mathrm{e}^{v_x}\right)\,dt
-v_x\,dx
\right.
\nonumber
\\
\fl
&&
\left.
+\left(u_x-\mathrm{e}^{v_x}\right)\,dy
\right).
\label{tilde_tilde_theta_20_Gerdjikov_eq}
\end{eqnarray}
This form is a solution to the equation
$d\omega = 2\,\alpha \wedge \omega +\Psi$,
where
$\Psi = \Phi +2\,\alpha\wedge \beta +d_{-2\alpha}(\theta_{0,0}-\theta_{1,0})$
is $d_{-2\alpha}$--cohomologous to the cocycle $\Phi +2\,\alpha\wedge \beta$.
Form  (\ref{tilde_tilde_theta_20_Gerdjikov_eq})  generates the Lax representation
\[
\left\{
\begin{array}{lcl}
v_t &=& \textfrac{1}{2}\,\left(\mathrm{e}^{v_x}-u_x\right)^2-u_y,
\\
v_y &=& \mathrm{e}^{v_x}-u_x
\end{array}
\right.
\]
for (\ref{Gerdjikov_eq}). This system was derived in \cite{Pavlov2003}.
\hfill$\diamond$

\vskip 10 pt

\noindent
\noindent
\addtocounter{example_counter}{1}
{\sc Example}
\arabic{example_counter}.
In the same way, for the Lie algebra
$\mathfrak{H}(4, 3)$
we have
\[
H^2_{\lambda\,\alpha} (\mathfrak{H}(4, 3)) =
\left\{
\begin{array}{lcl}
\mathbb{R}\,[\alpha \wedge \beta], & ~~~~& \lambda = -3,
\\
\{0\}, & ~~~~& \lambda \neq -3,
\end{array}
\right.
\]
and
\[
\beta = p^3\,dx,
\]
\[
\theta_{0,0} = \textfrac{1}{3}\,q^3\,dt,
\]
\[
\theta_{1,0} = -\textfrac{1}{2}\,p\,q^2\,\left(dy - 2\,u_x\,dt\right),
\]
\[
\theta_{2,0} = p^2\,q\,\left(dz-u_x\,dy-(u_z-u_x^2)\,dt\right),
\]
\[
\theta_{4,0} = p^4\,q^{-1}\,(du -u_t\,dt -u_x\,dx-u_y\,dy-u_z\,dz).
\]
Instead of equation for $d\theta_{3,0}$ from (\ref{structure_eqns_of_H_n_epsilon_r}) we consider equation
\[
\fl
d\tilde{\theta}_{3,0} = 3\,\alpha \wedge \tilde{\theta}_{3,0}
+\theta_{3,1}\wedge \theta_{0,0}
+\textfrac{2}{3}\,\theta_{2,1}\wedge \theta_{1,0}
+\textfrac{1}{3}\,\theta_{1,1}\wedge \theta_{2,0}
+\beta\wedge \theta_{0,1}
+3\,\alpha \wedge \beta.
\]
Then for $q=p\,\mathrm{exp}(v_x)$ we get the Wahlquist--Estabrook form
\[
\tilde{\theta}_{3,0}-\theta_{0,0}+\theta_{1,0}-\theta_{2,0} =
p^3\,\left(dv
-\left(\textfrac{1}{3}\,\left(\mathrm{e}^{v_x}-u_x\right)^3 - u_z\,\left(\mathrm{e}^{v_x} -u_x\right)
-u_y\right)\,dt
\right.
\]
\[
\qquad\qquad
\left.
-v_x\,dx
-\left(\textfrac{1}{2}\,\left(\mathrm{e}^{v_x}-u_x\right)^2-u_z\right)\,dy
-\left(\mathrm{e}^{v_x}-u_x\right)\,dz
\right)
\]
for the Lax representation
\[
\left\{
\begin{array}{lcl}
v_t &=& \textfrac{1}{3}\,\left(\mathrm{e}^{v_x}-u_x\right)^3 - u_z\,\left(\mathrm{e}^{v_x} -u_x\right)-u_y,
\phantom{\frac{\frac{A}{A}}{\frac{A}{A}}}
\\
v_y &=& \textfrac{1}{2}\,\left(\mathrm{e}^{v_x}-u_x\right)^2-u_z,
\phantom{\frac{\frac{A}{A}}{\frac{A}{A}}}
\\
v_z &=& \mathrm{e}^{v_x}-u_x
\phantom{\frac{\frac{A}{A}}{\frac{A}{A}}}
\end{array}
\right.
\]
of the system
\[
\left\{
\begin{array}{lcl}
u_{yy} &=& u_{tz} + u_z\,u_{zz},
\\
u_{zz} &=& u_{xy} +u_x\,u_{xz},
\\
u_{yz} &=& u_{tx}+ u_x\,u_{xy}+u_z\,u_{xz}.
\end{array}
\right.
\]
The first and the second equations of this system differ from the potential Khokhlov--Zabolotskaya equation
(\ref{pKhZ}) and equation (\ref{Gerdjikov_eq}), respectively, by notation.
\hfill$\diamond$

\vskip 10 pt

We suppose that for algebras $\mathfrak{H}(r+1,r)$ the same computations
will give higher elements of the integrable hierarchy generated by equation (\ref{Gerdjikov_eq}).

%%%%%%%%%%%%%%%%%%%%%%%%%%%%%%%%%%%%%%%%%%%%%%%%%%%%%%%%%%%%%%%%%%%%%%%%%%%%%%%%%%%%%%%%%%%%%%%%%%%%%%

\section{Second exotic cohomology group of $\mathfrak{G}(n, \varepsilon)$}

To obtain further generalizations of the constructions from Section \ref{section5}  we study
the second exotic cohomology group of the Lie algebra $\mathfrak{G}(n,\varepsilon)$.

\vskip 5 pt
\noindent
{\sc Theorem} 2. {\it  Suppose $n \ge 2$, then for each fixed $r \in \{2, \dots, n\}$
\[
H^2_{-r\,\alpha} (\mathfrak{G}(n, -\sfrac{2}{r})) = \mathbb{R}\,[\Phi_{r}],
\]
where
\[
\Phi_{r} = \sum \limits_{m=0}^{[r/2]} \left(r-2\,m\right)\,\theta_{r-m,0}\wedge\theta_{m,0}.
\]
For other values of $\lambda$ and $\varepsilon$ we have}
\[
H^2_{\lambda\,\alpha} (\mathfrak{G}(n,\varepsilon)) = \{0\}.
\]
\vskip 7pt
\noindent
Proof.
For forms $\eta$, $\omega_1$, ... , $\omega_k$ such that $\mathrm{deg}\, \eta \ge \mathrm{deg}\, \omega_i$
denote by $\eta \vert_{\omega_1=0, \dots, \omega_k=0}$ the result of re\-pla\-ce\-ment of each entry of
$\omega_1$, ... , $\omega_k$ in $\eta$ by $0$.
The condition $\eta = \eta \vert_{\omega_1=0, \dots, \omega_k=0}$ will be denoted as %by
$\eta = o(\omega_1, \dots, \omega_k)$.

From (\ref{structure_eqns_of_G_n_epsilon}) it follows that for each $k\ge 0$, $m\ge 0$
\[
d\theta_{k,m} = (k\,\alpha+(k\,\varepsilon +1 -m)\,\theta_{0,1}) \wedge \theta_{k,m} +o(\alpha, \theta_{0,1}),
\]
hence for $U=-\varepsilon Y+X_{0,1} \in \mathfrak{G}(n,\varepsilon)$ we have
\[
[U, X_{k,m}] = (m-1)\,X_{k,m}.
\]
Therefore $U$ defines an inner grading in $\mathfrak{G}(n,\varepsilon)$, \cite[\S~1.5.2]{Fuks1984}, that is,
\[
\mathfrak{G}(n,\varepsilon) = \bigoplus_{j=-1}^{\infty} \mathfrak{g}_j,
\qquad [\mathfrak{g}_i, \mathfrak{g}_j] \subseteq  \mathfrak{g}_{i+j},
\]
where $\mathfrak{g}_j = \left\{X \in \mathfrak{G}(n,\varepsilon)\,\,\vert\,\,[U,X]=j\,X \right\}$
for $j \ge -1$ and $\mathfrak{g}_j = \{0\}$ for $j <-1$.
We have $\mathfrak{g}_0 = \langle Y, X_{0,1} \rangle$ and
$\mathfrak{g}_j = \langle X_{k,j+1} \,\,\vert\,\,0 \le k \le n \rangle$ for $j \in \mathbb{N} \cup \{-1\}$, so
$\mathrm{dim}\,\mathfrak{g}_j \le n$.
From Remark 1
it follows that $H^{*}_{\lambda\alpha}(\mathfrak{G}(n,\varepsilon))$ is
the cohomology of $\mathfrak{G}(n,\varepsilon)$ with coefficients in the module
$A = \bigcup \limits_{\mu \in \mathbb{R}} A_\mu$,
where
\[
\fl
A_\mu=\left\{
a \in \mathbb{R}\,\,\vert\,\, \lambda\,\alpha(U)\cdot a = \mu \cdot a
\right\}
=
\left\{
a \in \mathbb{R}\,\,\vert\,\, -\varepsilon\,\lambda\,a = \mu \, a
\right\}
=
\left\{
\begin{array}{cll}
\mathbb{R}, &~~~& \mu= \varepsilon\,\lambda,
\\
\{0\}, &&  \mu \neq \varepsilon\,\lambda.
\end{array}
\right.
\]
For $m\ge 0$ and $\nu \in \mathbb{R}$ denote
\[
\fl
C_{(\nu)}^m (\mathfrak{G}(n,\varepsilon),A) = \left\{
\omega \in C^m (\mathfrak{G}(n,\varepsilon),A)\,\,\vert\,\,
\omega(W_1, \dots, W_m) \in A_{j_1+...+j_m-\nu} \,\,
\mathrm{for} \,\, W_k \in \mathfrak{g}_{j_k}
\right\}.
\]
In particular,
\begin{equation}
C_{(0)}^2 (\mathfrak{G}(n,\varepsilon),A) =
\bigoplus \limits_{
\begin{array}{c}
i+j = - \varepsilon\lambda,
\\
i \ge -1, j \ge -1
\end{array}
}
\mathrm{Hom}(\mathfrak{g}_i \wedge \mathfrak{g}_j, \mathbb{R}).
\label{C_0_2_G_n_e_A}
\end{equation}
In accordance with  \cite[Th.~1.5.2a]{Fuks1984} the inclusion
$C_{(0)}^{*}(\mathfrak{G}(n,\varepsilon),A) \rightarrow C^{*}(\mathfrak{G}(n,\varepsilon),A)$
in\-du\-ces a cohomological isomorphism. Hence, from (\ref{C_0_2_G_n_e_A}) it follows that
$H^2_{\lambda \alpha} (\mathfrak{G}(n,\varepsilon))$ is finite-dimensional for each fixed pair $\lambda$,
$\varepsilon$, and for an arbitrary solution $\Omega$ of equation $d\Omega +\lambda \,\alpha \wedge \Omega =0$
we can assume without loss of generality that
$M = \mathrm{min}\,\{j \in \mathbb{N} \cup \{0\}\,\,\vert\,\,
\Omega = o (\theta_{0,j+1}, \theta_{1,j+1}, \dots, \theta_{n,j+1})\,
\} <\infty$.
Then $\Omega = \sum \limits_{j=0}^n \gamma_j \wedge \theta_{j,M}$
with $\gamma_j = o(\theta_{j,M}, \theta_{j+1,M}, \dots \theta_{n,M})$.
Suppose $M \ge 1$.
From the structure equations (\ref{structure_eqns_of_G_n_epsilon}) it follows that
\[
d\theta_{k,M} = \sum \limits_{m=0}^k (1+\varepsilon\,m)\,\theta_{k-m,M+1} \wedge \theta_{m,0} +
o(\theta_{0,M+1}, \theta_{1,M+1}, \dots \theta_{n,M+1}).
\]
Hence $d\Omega +\lambda \,\alpha \wedge \Omega = \gamma_n \wedge \theta_{n,M+1} \wedge \theta_{0,0} +
o(\theta_{n,M+1}) = 0$ yields $\gamma_n = c\,\theta_{0,0}$ with $c\in \mathbb{R}$.
Thus $\Omega - d_{\lambda\,\alpha}(c\,\theta_{n,M-1}) =o(\theta_{n,M})$. In other words, we can put
$\gamma_n=0$ without loss of generality. In the same way we can consequently put $\gamma_{n-1}=0$, then
$\gamma_{n-2}=0$, etc., $\gamma_{0}=0$. So we have $M=0$ and
\[
\Omega = \sum \limits_{0 \le i < j \le n} a_{ij}\, \theta_{i,0} \wedge \theta_{j,0}+
\sum \limits_{j=0}^n b_j\,\alpha \wedge \theta_{j,0}
\]
with $a_{ij}, b_j \in \mathbb{R}$.

Then we have
$d\Omega+\lambda \,\alpha \wedge \Omega = - b_j\,\alpha\wedge \theta_{j,1} \wedge \theta_{0,0}
+o(\alpha\wedge \theta_{j,1} \wedge \theta_{0,0})$ for each $j \in \{0,1,\dots,n\}$,
so $b_0 =b_1 = \dots = b_n=0$. We assume that $a_{ij} \neq 0$ for some $i$, $j$, then
equation $d\Omega+\lambda \,\alpha \wedge \Omega
= a_{ij}\,(i+j+\lambda)\,\alpha\wedge \theta_{i,0} \wedge \theta_{j,0}+o(\theta_{i,0} \wedge \theta_{j,0})$
yields $i+j =r$, where $r \in \mathbb{N}$ is a constant, and $\lambda = -r$.
Therefore, $\Omega = \sum \limits_{j =0}^{[r/2]} c_j \theta_{j,0} \wedge \theta_{r-j,0}$ for some
$c_j \in \mathbb{R}$.

We claim that $r \le n$. Indeed, assume $r>n$ and put
$r_0 = \mathrm{min}\,\{j \in \mathbb{N} \,\,\vert\,\,\Omega =o(\theta_{j+1,0})\}$.
Since $r_0 \le n$, we have $r-r_0 \ge 1$. Then from
$\Omega = c_{r-r_0}\, \theta_{r-r_0,0} \wedge \theta_{r_0,0}+o(\theta_{r_0,0})$
we have $c_{r-r_0} \neq 0$. But
$d\Omega -r\,\alpha \wedge \Omega
= -c_{r-r_0}\,\theta_{r-r_0,0} \wedge \theta_{r_0,1} \wedge \theta_{0,0} + o(\theta_{r_0,1})$
implies $c_{r-r_0} = 0$. The contradiction proves the claim.

Consider the case of even $r$ and put $r=2\,m$ for $m\in \mathbb{N}$. Then
$\Omega = \sum \limits_{j =0}^{m-1} c_j \theta_{j,0} \wedge \theta_{2\,m-j,0}$ and
\begin{eqnarray*}
d\Omega-r\,\alpha \wedge \Omega &=&
\sum \limits_{j=1}^{m-1} \left(c_j-c_0\,(1+\varepsilon\,j)\right)\,
\theta_{0,0} \wedge \theta_{r-j,1}\wedge \theta_{j,0}
\\
&&+o(\theta_{0,0} \wedge \theta_{m+1,1}, \theta_{0,0} \wedge \theta_{m+2,1},
\dots, \theta_{0,0} \wedge \theta_{2m,1}
 ).
\end{eqnarray*}
So we get $c_j = c_0\,(1+\varepsilon\,j)$ and thus
\begin{equation}
\Omega = c_0\,\sum \limits_{j =0}^{m-1} (1+\varepsilon\,j)\, \theta_{j,0} \wedge \theta_{r-j,0}.
\label{Omega_r}
\end{equation}
Finally we have
\[
d\Omega-r\,\alpha \wedge \Omega = (1+\varepsilon\,m)\,
\left(
d\theta_m\wedge \theta_m+2\,\sum\limits_{j=0}^{m-1}\,d\theta_{j,0} \wedge \theta_{2m-j,0}
\right),
\]
and hence $\varepsilon = -\frac{1}{m} = -\frac{2}{r}$. Substituting this into
(\ref{Omega_r}) with $c_0=-r$ yields $\Omega=\Phi_r$.

In the case $r=2\,m+1$ the proof is similar.
\hfill $\square$ %{\sc q.e.d.}

\vskip 10 pt
\noindent
{\sc Corollary}.
{\it For each $r\in \{2, \dots, n\}$ the equation
\begin{equation}
d\omega =r\,\alpha \wedge \omega+\Phi_{r}.
\label{d_omega_Phi_r}
\end{equation}
is compatible with the structure equations of} $\mathfrak{G}(n, -\sfrac{2}{r})$.

\vskip 10 pt
\noindent
\noindent
\addtocounter{example_counter}{1}
{\sc Example}
\arabic{example_counter}.
In the case of $n=r=3$ integration of the structure equations of
$\mathfrak{G}(3, \sfrac{-2}{3})$ gives
\[
\theta_{0,0} = -\textfrac{1}{3}\,q^3\,dt,
\]
\[
\theta_{1,0} = -p\,q\,\left(dy +u_x\,dt\right),
\]
\[
\theta_{2,0} = p^2\,q^{-1}\,\left(dx-u_x\,dy-(u_y+u_x^2)\,dt\right),
\]
\[
\theta_{3,0} = p^3\,q^{-2}\,(du -u_t\,dt -u_x\,dx-u_y\,dy),
\]
while equation (\ref{d_omega_Phi_r}) acquires the form
\[
d\omega = 3\,\alpha \wedge \omega
+3\,\theta_{3,0}\wedge\theta_{0,0}
+\theta_{2,0}\wedge\theta_{1,0}.
\]
We integrate this to obtain the form $\omega$, then rename
$q=p^{-1}\,v_x^{-1}$ and consider
\[
\omega+\theta_{2,0} = p^3\,\left(
dv-(u-(u_x^2+u_y)\,v_x)\,dt-v_x\,dx-(x-u_x\,v_x)\,dy
\right).
\]
This form defines the Lax representation
\[
\left\{
\begin{array}{lcl}
v_t &=& u-(u_x^2+u_y)\,v_x,
\\
v_y &=& x-u_x\,v_x
\end{array}
\right.
\]
for the equation
\[
u_{yy} = u_{tx}+(u_y-u_x^2)\,u_{xx}-3\,u_x\,u_{xy}.
\]
This Lax representation was found in \cite{Morozov2009}.
\hfill$\diamond$

\vskip 10 pt
\noindent
\noindent
\addtocounter{example_counter}{1}
{\sc Example}
\arabic{example_counter}.
In the case $n=r=4$ instead of
$\mathfrak{G}(4,\sfrac{-1}{2})$ we consider the Lie algebra
$\mathfrak{H}(4,2)$. We have
\[
H^2_{\lambda\,\alpha} (\mathfrak{H}(4,2)) =
\left\{
\begin{array}{lcl}
\mathbb{R}\,[\alpha \wedge \beta], & ~~~~& \lambda = -2,
\\
\mathbb{R}\,[\tilde{\Phi}_4], & ~~~~& \lambda = -4,
\\
\{0\}, & ~~~~& \lambda \not \in \{ -2, -4\},
\end{array}
\right.
\]
where
\[
\tilde{\Phi}_4 =\frac{1}{2}\,\Phi_4+ \beta \wedge \theta_{2,0}
= 2\,\theta_{4,0} \wedge \theta_{0,0}+\theta_{3,0}\wedge \theta_{1,0} + \beta\wedge\theta_{2,0}.
\]
Consider the case $\lambda = -4$. From the structure equations of $\mathfrak{H}(4,2)$ we obtain
\[
\beta = \textfrac{1}{2}\,p^2\,dx,
\]
\[
\theta_{0,0} = -\textfrac{1}{2}\,q^2\,dt,
\]
\[
\theta_{1,0} = p\,q\,\left(dy -u_z\,dt\right),
\]
\[
\theta_{2,0} = p^2\,\left(dw-(u_x-u_z^2)\,dt-\ln \vert q \vert \,dx+u_z\,dy\right),
\]
\[
\theta_{3,0} = p^3\,q^{-1}\,\left(dz+\left(u_y-u_x u_z-\textfrac{1}{2}\,u_z^3\right)\,dt
-u_z\,dx+\left(u_x+\textfrac{1}{2}\,u_z\right)\,dy\right),
\]
\[
\theta_{4,0} = p^4\,q^{-2}\,(du -u_t\,dt -u_x\,dx-u_y\,dy-u_z\,dz).
\]
Then we integrate the equation
\[
d\omega = 4\,\alpha \wedge \omega+\tilde{\Phi}_4
\]
and get
\[
\omega = p^4\,\left(dv-u\,dt+z\,dy-w\,dx\right).
\]
After the change of notation $q=p^{-1}\,v_z^{-1}$, $w = v_x+u_z\,v_z$ we have
\[
%\fl
\omega -\theta_{3,0}= p^4\,\left(
dv
-\left(u+\left(u_y-u_x\,u_z-\textfrac{1}{2}\,u_z^3\right)\,v_z\right)\,dt
-v_x\,dx
\right.
\]
\[
\qquad
\left.
-\left(\left(u_x+\textfrac{1}{2}\,u_z\right)\,v_z-z\right)\,dy
-v_z\,dz
\right).
\]
This is the Wahlquist--Estabrook form for the Lax representation
\[
\left\{
\begin{array}{lcl}
v_t &=& u+\left(u_y-u_x\,u_z-\textfrac{1}{2}\,u_z^3\right)\,v_z,
\\
v_y &=& z-\left(u_x+\textfrac{1}{2}\,u_z\right)\,v_z.
\end{array}
\right.
\]
of the equation
\[
%\fl
u_{yy} = u_{tx}+u_z\,u_{xy}-u_y\,u_{xz} +u_z\,u_{tz}+2\,(u_x+u_z^2)\,u_{yz}
\]
\[
\qquad
-
\left(u_x^2+u_x\,u_z^2+u_y\,u_z+\textfrac{1}{4}\,u_z^4\right)\,u_{zz}.
\]
\hfill$\diamond$

%%%%%%%%%%%%%%%%%%%%%%%%%%%%%%%%%%%%%%%%%%%%%%%%%%%%%%%%%%%%%%%%%%%%%%%%%%%%%%%%%%%%%%%%%%%%%%%%%%%%%%

\section{Concluding remarks}

In the present paper we have shown that both known (Examples 2, 3, 4, and 6) and new (Examples 5 and 7) Lax
representations for integrable {\sc pde}s can be derived from the second exotic cohomology group of
symmetry pseudo-groups of the {\sc pde}s under the study. This approach gives the solution to the problem of
existence of a Lax representation in internal terms of the {\sc pde} and allows one to eliminate
apriori assumptions about the possible form of the Lax representation. The approach is universal
and can be used to analyze a lot of equations or Lie algebras with nontrivial second exotic cohomology group.
Quite natural this leads to the question of generalization of the Lie algebras considered above. In particular,
it seems to be important to study the adjoint cohomology group
$H^1(\mathfrak{G}(n,\varepsilon),\mathfrak{G}(n,\varepsilon))$
in order to describe right extensions of $\mathfrak{G}(n,\varepsilon)$ and to generalize the Lie algebras
$\mathfrak{H}(n,r)$. Another challenge is to replace the Lie algebra of vector fields on $\mathbb{R}$ by
Lie algebras of vector fields on $\mathbb{R}^n$ in the constructions of  $\mathfrak{G}(n,\varepsilon)$
and $\mathfrak{H}(n,r)$, to consider exotic cohomology of the resulting Lie algebras and to
study whether there exist related integrable systems. For example, the symmetry pseudo-groups
of the heavenly equations and their deformations are right extensions of Lie algebras of the form
$\EuScript{A} \otimes \mathfrak{h}$, where $\mathfrak{h}$ is the Lie algebra of Hamiltonian vector fields on
$\mathbb{R}^2$ and $\EuScript{A}$ are algebras of truncated polynomials,
\cite{KruglikovMorozov2012,KruglikovMorozov2015}. Other interersting examples include equations discussed in
\cite{KruglikovMorozov2016}. It is natural to elucidate the relationship
among the Lax representations and the structure of symmetry pseudogroups for the above-mentioned integrable
equations. We intend to address this issue in our future work.

\section*{Acknowledgments}
This research was supported in part by the Polish Ministry of Science and Higher Education.
I am  grateful to the organizers of the conference ``Geometry and Algebra of PDEs''
held at  the Arctic University of Norway, Troms{\o}, 6--10 June 2017,  for financial sup\-port.

I am pleased to thank
V.V. Lychagin, I.S. Krasil${}^{\prime}$shchik, P.J. Olver, B.S. Kruglikov,
D.V. Alekseevsky, B.A. Khesin and P. Zusmanovich for useful and
stimulating discussions, and
to V.S. Gerdjikov for pointing out the reference \cite{Gerdjikov1990} and providing the copy thereof.


\section*{References}
\begin{thebibliography}{99}

\bibitem{Blaszak2002} M. B{\l }aszak.
   Classical R-matrices on Poisson algebras and related dispersionless systems.
   Phys. Lett. A \textbf{297}   (2002), 191--195

\bibitem{VinogradovKrasilshchik1997} A.V. Bocharov et al.,
 {\it Symmetries of Differential Equations in Mathematical Physics and
 Natural Sciences}, edited by A.M. Vinogradov  and I.S. Krasil$^\prime$shchik.
  Factorial Publ.\ House, 1997 (in  Russian). English translation: Amer.\ Math.\ Soc., 1999

\bibitem{CarollKodama1995}  R. Carroll. Y. Kodama.
 Solution of the dispersionless Hirota equations.
 J. Phys. A, {\bf 28} (1995), 6373--6387

\bibitem{Cartan1} \'E. Cartan,
    Sur la structure des groupes infinis de transformations,
    {\it {\OE}uvres Compl{\`e}tes},  Part II,  V. 2, 571--714,
    Paris, Gauthier - Villars, 1953.

\bibitem{Cartan4} \'E. Cartan,
    La structure des groupes infinis,
    {\it {\OE}uvres Compl{\`e}tes}, Part II,  V. 2, 1335--1384,
    Paris, Gauthier - Villars, 1953.

\bibitem{DubrovinNovikov1989}
B.A. Dubrovin, S.P. Novikov. Hydrodynamics of weakly deformed soliton lattices.
Differential geometry and Hamiltonian theory.
Russian Mathematical Surveys, {\bf 44}:6 (1989), 35--124

\bibitem{LNM810}
{\it Geometrical Approaches to Differential Equations}, R. Martini (Ed.)  Lecture Notes in Mathematics,
Springer, 1980

\bibitem{Gerdjikov1990}
V. S. Gerdjikov. $Z_N$--reductions and new integrable versions of derivative nonlinear Schr{\"o}dinger
equations. In: {\it Nonlinear Evolution Equations: Integrability and Spectral Methods}, Ed. A. P. Fordy,
A. Degasperis, M. Lakshmanan, Manchester University Press, 1990, 367--372.

\bibitem{Gibbons1984} J. Gibbons. The Zabolotskaya--Khokhlov equation and the inverse scattering problem
    of classical mechanics. In:   {\it Dynamical Problems in Soliton Systems}, Ed. S. Takeno, Springer 1985,
    36--41

\bibitem{GibbonsTsarev1996}
J. Gibbons, S.P. Tsarev. Reductions of the Benney equations.
Phys. Lett. A, {\bf 211} (1996), 19--24

\bibitem{Fuks1984} D.B. Fuks. {\it Cohomology of Infinite-Dimensional Lie Algebras}.
Consultants Bureau, N.Y., 1986

\bibitem{IveyLandsberg} T.A. Ivey, J.M. Landsberg.
{\it Cartan for Beginners: Differential Geometry via Moving Frames and Exterior Differential
Systems}. AMS, Providence, Rhode Island, 2003

\bibitem{Kodama1988} Y. Kodama.
A method for solving the dispersionless KP equation and its exact solutions.
Phys. Lett. A
{\bf 129} (1988), 223--226

\bibitem{KrasilshchikVinogradov1984} I.S. Krasil${}^{\prime}$shchik, A.M. Vinogradov, Nonlocal symmetries and the theory of
    coverings. Acta Appl. Math. {\bf 2} (1984) 79--86

\bibitem{KrasilshchikVinogradov1989} I.S. Krasil$^\prime$shchik,
A.M. Vinogradov. Nonlocal trends in the geometry of differential equations: symmetries, conservation laws,
and B\"{a}cklund transformations.
Acta Appl.\ Math. \textbf{15} (1989),
161--209

\bibitem{KrasilshchikVerbovetsky2011} I.S. Krasil${}^{\prime}$shchik, A.M. Verbovetsky, R. Vitolo,
    A unified approach to computation of integrable structures,
    Acta Appl. Math. {\bf 120} (2012)  199--218

\bibitem{KrasilshchikVerbovetskyVitolo2012} J. Krasil${}^{\prime}$shchik, A. Verbovetsky,
    Geometry of jet spaces and integrable systems,
    J. Geom. Phys. {\bf 61} (2011) 1633--1674

\bibitem{Krichever1990} I.M. Krichever. Method of averaging for two-dimensional "integrable" equations.
 Functional Analysis and Its Applications {\bf 22}:3 (1988) 200--213


\bibitem{KruglikovMorozov2012}
B.S. Kruglikov, O.I. Morozov. SDiff(2) and uniqueness of the Pleba{\'n}ski equation.
J. Math. Phys. {\bf 53} (2012), 083506, 11 pp.

\bibitem{KruglikovMorozov2015}
B.S. Kruglikov, O.I. Morozov. Integrable dispersionless PDEs in 4D, their symmetry pseudogroups and
deformations. Lett. Math. Phys. {\bf 105} (2015) 1703--1723

\bibitem{KruglikovMorozov2016}
B.S. Kruglikov, O.I. Morozov.
A B\"acklund transformation between the four-dimensional Mart{\'{\i}}nez Alonso--Shabat
and Ferapontov--Khus\-nut\-di\-no\-va equations.
Theor. Math. Phys. {\bf 188}(3) (2016) 1358--1360

\bibitem{Kupershmidt1990} B.A. Kupershmidt.
The quasiclassical limit of the modified KP hierarchy.
J. Phys. A, {\bf 23} (1990), 871--886


\bibitem{Kuzmina1967} G.M. Kuz${}^{\prime}$mina, On a possibility to reduce a system of two first-order partial
        differential equations to a single equation of the second order,  Proc. Moscow State
        Pe\-da\-go\-gi\-cal Institute {\bf 271} (1967), 67--76 (in Russian).

\bibitem{LinReissnerTsien} C.C. Lin, E. Reissner, H.S. Tsien,
    On two-dimensional non-steady motion of a slender body in a compressible fluid,
    J. Math. Phys. {\bf 27} (1948) 220--231



\bibitem{Morozov2009}  O.I. Morozov.
 Contact integrable extensions of symmetry pseudo-groups and
coverings of (2 + 1) dispersionless integrable equations.
J. Geom. Phys. \textbf{59} (2009), 1461--1475

\bibitem{Morozov2017}
O.I. Morozov. Deformed cohomologies of symmetry pseudo-groups
and coverings of differential equations. J. Geom. Phys. {\bf 113} (2017), 215--225

\bibitem{Novikov2002} S.P. Novikov. On the exotic De-Rham cohomology. Perturbation theory as a spectral sequence.
{\tt arXiv:math-ph/0201019}

\bibitem{Olver1995} P.J. Olver.
{\it Equivalence, Invariants, and Symmetry}. Cambridge University Press, Cambridge, 1995.

\bibitem{Pavlov2003} M.V. Pavlov. Modified dispersionless Veselov--Novikov equations and
corresponding hydrodynamic chains. {\tt arXiv:nlin/0611022}

\bibitem{Stormark2000} O. Stormark, {\it Lie's Structural Approach to PDE Systems}, Encyclopedia Math. Appl.,
    80, Cam\-brid\-ge University Press, Cambridge, 2000

\bibitem{Vasilieva1972} M.V. Vasil${}^{\prime}$eva, {\it Structure of Infinite Lie Groups of Transformations},
    Moscow State Pe\-da\-go\-gi\-cal Institute, Moscow, 1972 (in Russian)

\bibitem{KhZ} E.A. Zabolotskaya, R.V. Khokhlov, Quasi-plane waves in the nonlinear acoustics
     of confined beams, Sov. Phys. Acoust. {\bf 15} (1969) 35--40

\bibitem{Zakharov1980} V.E. Zakharov. Benney equations and quasiclassical approximation
in the method of the inverse problem.
Functional Analysis and Its Applications, 1980, 14:2, 89–98

\bibitem{Zakharov1991}
{\it What is integrability}, V.E. Zakharov (Ed.)  Springer Series in
Nonlinear Dynamics, Springer, 1991

\bibitem{Zakharov1994}
V.E. Zakharov, Dispersionless limit of integrable systems in 2+1 dimensions, in:
{\it Singular Limits of Dispersive
Waves}, Ed. N.M. Ercolani et al., Plenum Press, NY (1994) 165--174


\end{thebibliography}
\end{document}